\documentclass[nocite,dvips]{epl}

\title{The penetration depth in Sr$_2$RuO$_4$:\\ Evidence for orbital
  dependent superconductivity}
\shorttitle{The penetration depth in Sr$_2$RuO$_4$}
\author{Hiroaki Kusunose \and Manfred Sigrist}
\institute{
Institut f\"ur Theoretische Physik - ETH-H\"onggerberg,
  CH-8093 Z\"urich, Switzerland
}

\begin{document}

\maketitle

\begin{abstract}
The apparent $T^2$-temperature dependence of the London penetration depth in
Sr$_2$RuO$_4$ is discussed on the basis of a multi-gap model with
horizontal line nodes. The influence of the nodes in combination with
nonlocal electromagnetic response leads to low-temperature behaviors as
$-T^2\ln T$ in the London limit and as $T^2$ in the Pippard limit.
These behaviors appear only at very low temperature.
On the other hand, the interplay of the superconductivity of
different bands is responsible for the observed $T^2$-like
behavior over the wide temperature range. The experimental data can
be fitted well with a set of material parameters as was used in the specific heat fitting.
\end{abstract}

Since the discovery of the superconductivity in the layered perovskite
compound Sr$_2$RuO$_4$ \cite{Maeno94}, much effort has been devoted to
the identification of its Cooper pairing symmetry \cite{Maeno01}.
An early theoretical prediction of the spin-triplet pairing
\cite{Rice95} is supported by the NMR Knight shift measurement, which
shows no change in the spin susceptibility on passing through the
superconducting transition at $T_c$ (=1.5K) \cite{Ishida98}.
Moreover,  $\mu$SR measurements reveal the appearance of
an intrinsic magnetism with the onset of superconductivity, which was
attributed to pairing with broken time reversal symmetry \cite{Luke98}.
These two experiments consistently identify the superconducting phase
as the chiral p-wave state $ {\bm d}_1({\bm k})={\bm
  z}\Delta_1(T)(\hat{k_x}+i\hat{k_y})$,
an analog to the A-phase of the superfluid $^3$He.
Here a representative form of the gap function,
$\Delta_{\alpha\beta}({\bm
  k})=(i\sigma^y{\bm\sigma})_{\alpha\beta}\cdot{\bm d}({\bm k})$, is
given in the $d$-vector representation \cite{Rice95,Hasegawa00}.
Such a gap function arises in a weak-coupling theory for a
two-dimensional Fermi liquid with spin-orbit coupling
\cite{Rice95,Sigrist99,Mackenzie96}.
This pairing state suggests a nodeless quasiparticle gap,
$\Delta_1(T)$.

As sample quality improved, however, power-law temperature
dependence was observed in various quantities, such as specific
heat \cite{NishiZaki9900},
NQR relaxation rate \cite{Ishida00}, thermal conductivity
\cite{Tanatar01,Izawa01} and ultrasonic attenuation
\cite{Lupien01}.
These findings seemed to indicate the unexpected existence of line nodes in
the gap and motivated the investigation of models with vertical
(pseudo) line nodes \cite{Miyake99,Graf00,Dahm00}.
Heat transport measurements, however, gave no evidence of strong
anisotropy of the gap in the basal plane, essentially excluding
vertical line nodes \cite{Tanatar01,Izawa01}.
If at all, nodes would have to be horizontal, i.e., parallel to the
$k_x$-$k_y$-plane \cite{Hasegawa00}. Still such a gap structure alone
seemed insufficient to account for all the thermodynamic
data.

Motivated by these experimental situations and the fact that Sr$_2$RuO$_4$ has
three conduction bands, called $ \alpha $, $ \beta $ and $\gamma$,
Zhitomirsky and Rice (ZR) proposed a multi-gap scenario, in which the primary nodeless
gap, ${\bm d}_1({\bm k})$, is developed in the single ``active'' band ($\gamma$-sheet),
while the secondary gaps with horizontal line nodes of the form,
\begin{equation}
{\bm d}_2({\bm k})={\bm z}\Delta_2(T)(\hat{k_x}+i\hat{k_y})\cos(k_z c/2),
\end{equation}
appear in the other (``passive'') bands ($\alpha$, $\beta$-sheets)
through an interband Cooper pair scattering \cite{Zhitomirsky01}.
Then this scenario was used to give a very good
fit in the specific heat over the whole temperature range.
The picture of an orbital dependent
superconductivity (ODS) \cite{Agterberg97} provides a consistent view
in many other aspects. Identifying the $\gamma$-band as the dominant
band for superconductivity makes sense in view of the fact that
ferromagnetic spin fluctuations, beneficial to spin-triplet pairing,
are slightly enhanced in this band \cite{Imai98,Ng01}, while the other two
bands show very strong spin correlations for the finite incommensurate wave
vector \cite{Mazin9799,Sidis99}. In addition, the chiral p-wave state
is stabilized by spin-orbit coupling relative to all other triplet
states, if the $ \gamma $-band is dominant \cite{Ng00}.

On general grounds, however, all scenarios involving line nodes are
hampered by the fact that line nodes can appear in a spin triplet
pairing state only accidentally, if spin-orbit coupling is
present. Blount has shown that, in general, only point nodes appear by
symmetry reasons \cite{Blount85}. In order to generate rigorous
line nodes the ZR scenario requires exclusively the interlayer pairing for
the $ \alpha $-$\beta$-bands. Any inplane pairing component would
leave point nodes only, in a rigorous sense. Nevertheless, we will
assume the presence of horizontal line nodes for the time being, and later
discuss consequences of the presence of the inplane pairing components.

In contrast to other thermodynamic quantities, the apparent
$T^2$-dependence over wide temperature range below 0.8 K observed in the London
penetration depth, $\lambda(T)$, for fields along the $z$-axis,
 \cite{Bonalde00} seems incompatible with line nodes, since a simple-minded
theory would lead a $T$-linear behavior. Bonalde {\it et al}. ascribed $T^2$-behavior
in their data to nonlocal effects, as was discussed by Kosztin and
Leggett (KL) in the context of the high-temperature superconductors \cite{Bonalde00,Kosztin97}.
The KL-theory is based on the fact that for the large limit of the
Ginzburg-Landau (GL) parameter $\kappa=\lambda(0)/\xi_0$ (London
limit), a small portion of the Fermi surface in the vicinity of the
nodes ($\sim \kappa^{-1}$) should be treated in the nonlocal limit.
This would alter the power-law from the linear to a $T^2$-behavior.
While this discussion is applicable rigorously for very large
$\kappa$ and in the very low-temperature regime only, Bonalde {\it et
  al}. argued that the relatively small $\kappa$ ($\approx 2$-$3$) of
Sr$_2$RuO$_4$ could lead to a wider temperature range, even up to $
T^*\sim\Delta(0)\kappa^{-1}\sim 0.6 T_c$, below which the KL renormalization appears.
Although there were several further theoretical explanations for the
peculiar power-law behavior \cite{Dahm00,Morinari00,Kubo00}, a clear
understanding has not been achieved yet.

In this Letter, we show that the rigorous asymptotic power-law
behaviors in $\lambda(T)$
due to nonlocal effects are restricted to very low temperatures
in both
the London ($\kappa\gg1$) and the Pippard ($\kappa\ll1$) limits.
Instead, based on the ODS model the overall $T^2$-dependence and beyond can be fitted well
and consistently with the specific heat data.
Thereby the non-locality of the electromagnetic response is
essential for the fitting of the experimental data and the resultant
Ginzburg-Landau (GL) parameter $\kappa=2.3$ is good agreement with
the experimental value.

We restrict our discussion to the clean limit and choose the sample
geometry so that the $x$-$z$-plane surface is exposed to magnetic
field parallel to the $z$-axis, so that
both the vector potential ${\bm A}$ and the
screening currents ${\bm j}$ are oriented parallel to the
$x$-axis.
The penetration depth is obtained from the electromagnetic
response kernel,
$K(q,T)=-(4\pi/c)j(q)/A(q)$, as
\begin{equation}
\lambda(T) = \frac{2}{\pi}\int_0^\infty \frac{dq}{q^2+K(q,T)},
\label{pene}
\end{equation}
assuming a specularly reflecting surface.

We only consider intraband pairing so that in the lowest order each
band separately contributes to the response kernel,
\begin{equation}
K(q,T)=\frac{1}{\lambda_0^2}\sum_l\zeta_l \tilde{K}_l(q,T).
\end{equation}
Here the zero-temperature penetration depth in the London limit has been defined
as $\lambda_0=\sqrt{mc^2/4\pi ne^2}$ with $n/m=\sum_l(n_l/m_l)$,
where $n_l$ and $m_l$ are the density and the effective mass of electrons in the
band $l$.
The ratio, $\zeta_l=(n_l/m_l)/(n/m)$, gives a measure for the
contribution of each band to the current density at $T=0$.
The dimensionless response kernel has the form
\begin{equation}
\tilde{K}_l(q,T)=2\pi T\sum_{n=-\infty}^\infty\sum_{k_z}
\frac{2}{\pi}\int_0^1d\mu
\frac{\sqrt{1-\mu^2}|{\bm d_l}|^2}{\sqrt{\omega_n^2+|{\bm
      d_l}|^2}(\omega_n^2+|{\bm d_l}|^2+Q_l^2\mu^2)},
\label{kernel}
\end{equation}
where $\omega_n=(2n+1)\pi T$ is the fermionic Matsubara frequency.
Here, we have assumed cylindrical Fermi surfaces for simplicity and
used $(\partial |{\bm d}_l|^2/\partial {\bm k})\cdot{\bm q}=0$ for the
present geometry.
The nonlocality appears via $Q_l=qv_{{\rm F}l}/2=(\pi\tilde{q}/2\kappa_{0l})\Delta_l(0)$
where the dimensionless variables $\tilde{q}=q\lambda_0$ and
$\kappa_{0l}=\lambda_0/\xi_{0l}$ with the coherence length for each
bands, $\xi_{0l}=v_{{\rm F}l}/\pi\Delta_l(0)$, were introduced.
For nodeless gaps or gaps with horizontal line-nodes, the
angular integral $\mu$ in eq.~(\ref{kernel}) yields
\begin{equation}
\tilde{K}_l(q,T)=\frac{2\pi
  T}{Q_l^3}\sum_{n=-\infty}^\infty\sum_{k_z}|{\bm
  d_l}|^2g\left(\frac{\sqrt{\omega_n^2+|{\bm d_l}|^2}}{Q_l}\right),
\label{kernel2}
\end{equation}
where $g(z)=(\sqrt{1+z^{-2}}-1)/z$.
At zero temperature, replacing $2\pi T\sum_n$ by
$\int_{-\infty}^\infty d\omega$, we obtain
\begin{equation}
\tilde{K}_l(q,0)=\sum_{k_z}g_0\left(\frac{Q_l}{|{\bm d}_l|}\right),
\label{kernel0}
\end{equation}
where $g_0(z)=(2/z)\tan^{-1}(z)-\ln(1+z^2)/z^2$.

We consider now the contributions from the bands with horizontal line
nodes, i.e., the passive bands in the ZR scenario ($l=2$), since their
quasiparticles would dominate the low-temperature behavior, while those
in the totally gapped dominant band are inactive.
Analogous to the KL theory, we evaluate the average over $k_z$ for an approximate
gap form in the vicinity of the nodes.
It leads to
\begin{equation}
\delta\tilde{K}_l(q,T)=\tilde{K}_l(q,T)-\tilde{K}_l(q,0)=
\delta\tilde{K}_l(0,T)F(\tilde{q}/t),
\label{dkernel}
\end{equation}
where $\delta\tilde{K}_l(0,T)=-2\ln2(T/\Delta_2(0))$ is the usual $T$-linear
contribution to the kernel and $t=T/T_{\rm L}$ with $T_{\rm
  L}=Q_2/\tilde{q}=(\pi/2\kappa_{02})\Delta_2(0)$.
For horizontal line nodes the universal function $F(z)$ has a form
different from the KL-expression:
\begin{equation}
F(z)=1-\frac{1}{\ln 2}\int_0^z dx f(x) \left(1-\frac{x}{z}\right)^2,
\end{equation}
and $f(x)=1/(e^x+1)$.
Note that $F(z)$ has the asymptotic form, $F(z)\approx \pi^2/6z\ln2$ for
$z\gg1$ and $1-z/6\ln2$ for $z\ll1$.
Since $\tilde{K}_2(q,0)=1-\pi\tilde{q}/6\kappa_{02}\approx1$ in the
London limit ($\kappa_{02}\gg1$) (eq.~(\ref{kernel0})), the deviation
of $\lambda(T)$ from its zero-temperature
value, $\lambda(0)$ ($=\lambda_0$), is given by
\begin{equation}
\frac{\Delta\lambda(T)}{\lambda_0}= \frac{\lambda(T) -
  \lambda(0)}{\lambda_0}= -\frac{2}{\pi}\int_0^\infty
d\tilde{q}\frac{\delta\tilde{K}_2(q,T)}{(\tilde{q}^2+1)^2}.
\label{pelon}
\end{equation}
An appropriate estimate of eq.~(\ref{pelon}) gives the following
asymptotic expression
for $T<0.05T_{\rm L}$,
\begin{equation}
\frac{\Delta\lambda(T)}{\lambda_0}=
-\frac{\pi^2}{3\kappa_{02}}\left(\frac{T}{T_{\rm
      L}}\right)^2\left(1.0+\ln\frac{T}{T_{\rm L}}\right).
\end{equation}

In the Pippard limit ($\kappa_{02}\ll1$) we use $g(z)=z^{-2}$ in
eq.~(\ref{kernel2}).
Then, the Matsubara sum can be evaluated straightforwardly
\begin{equation}
\tilde{K}_2(q,T)=\frac{\pi}{Q_2}\sum_{k_z}|{\bm
  d}_2|\tanh\left(\frac{|{\bm d}_2|}{2T}\right).
\label{pkernel}
\end{equation}
From eq.~(\ref{pene}) we obtain
\begin{equation}
\frac{\lambda(T)}{\lambda(0)}=\left[\frac{\pi}{2}\sum_{k_z}\frac{|{\bm
      d}_2|}{\Delta_2(0)}\tanh\left(\frac{|{\bm
        d}_2|}{2T}\right)\right]^{-1/3},
\end{equation}
where $\lambda(0)/\lambda_0=(2\pi/3\sqrt{3})(\pi/4\kappa_{02})^{1/3}$.
At low temperature the main contribution in eq.~(\ref{pkernel}) comes
from $k_z\sim \pm\pi/c$.
Evaluating the $k_z$-average in eq.~(\ref{pkernel}) or using the
large-$\tilde{q}$ limit of eqs.~(\ref{kernel0}) and (\ref{dkernel}),
the low-temperature expression is given by
\begin{equation}
\frac{\Delta\lambda(T)}{\lambda(0)}=\frac{1}{18}\left(\frac{T}{T_{\rm
      P}}\right)^2+\frac{83}{3240}\left(\frac{T}{T_{\rm P}}\right)^4,
\end{equation}
where $T_{\rm P}=\Delta_2(0)/\pi$.
The deviation from the $ T^2 $-dependence
becomes visible for $ T$ larger than $ 0.3 T_{\rm P}$. Note that the
prefactor of the $ T^2 $ term in both limits is
determined by the inverse of the characteristic temperatures, i.e. the
smaller $\Delta_2(0)$ gives the faster increase of $\lambda(T)$ close
to $ T= 0 $.


In conclusion, the power-law behaviors due to nonlocal effect
are restricted to temperature below $T^*$.
For any $\kappa$, $T^*$ has a value between both limits, i.e., $0.05T_{\rm L}<T^*<0.3T_{\rm P}$,
which is much lower than $\Delta_2(0)$.
Since $T^* < 0.17T_c$ even with the maximal gap, $\Delta_2(0)=\Delta_{\rm BCS}$ ($=1.76T_c$),
it already excludes a reasonable fitting of the $ T^2
$-behavior with a single-gap model including horizontal line
nodes for Sr$_2$RuO$_4$.

\begin{figure}[ht]
\onefigure{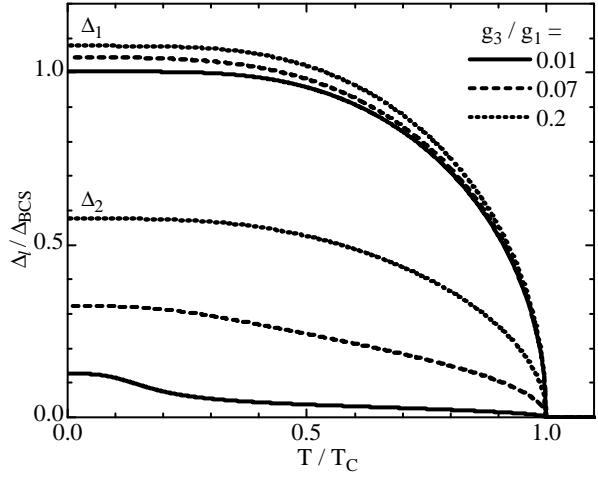}
\caption{\label{fig1} The temperature dependence of the gap functions
  scaled by the BCS gap at $T=0$. The interband couplings are varied,
  while the other parameters are fixed as $g_1=0.4$ and
  $g_2/g_1=0.85$.}
\end{figure}
We now turn to the multi-gap scenario for Sr$_2$RuO$_4$\cite{Zhitomirsky01},
in which the temperature dependence of the gap magnitudes are given by
an effective two-band model; $\Delta_\alpha=\Delta_\beta=\Delta_2(T)$ and
$\Delta_\gamma=\Delta_1(T)$.
In the gap equations, the pairing interactions are defined as
$V_{11}=-g_1\hat{\bm k}\cdot\hat{\bm k}'$, $V_{22}=-2g_2(\hat{\bm
  k}\cdot\hat{\bm k}')\cos(k_zc/2)\cos(k_z'c/2)$ for intraband coupling and
$V_{12}=-\sqrt{2}g_3(\hat{\bm k}\cdot\hat{\bm k}')\cos(k_z'c/2)$ for
the interband Cooper pair scattering.
We estimate various other material parameters from the de
Haas-van Alphen measurement
\cite{Mackenzie96}: the Fermi velocities, $v_{{\rm F}\alpha}\sim
v_{{\rm F}\beta}=0.68v_{{\rm F}\gamma}$, the (Fermi surface averaged)
band masses over their carrier densities,
$\zeta_\alpha:\zeta_\beta:\zeta_\gamma=0.20:0.44:0.36$ and the
densities of states at the Fermi level, $N_{01}:N_{02}=0.57:0.43$.
We fix the parameters as $g_1=0.4$ (in units of the inverse total
density of states), $g_2/g_1=0.85$ and
$\kappa_{02}/\kappa_{01}=(\Delta_2(0)/\Delta_1(0))/0.68$.
This leaves two free parameters for fitting: $g_3/g_1$ and
$\kappa_{01}$.

Figure~\ref{fig1} shows that the temperature dependence of the gaps
$\Delta_l(T)$ for different interband couplings $g_3/g_1$.
In the absence of interband coupling ($g_3 =0 $) a second phase
transition occurs for $ \Delta_2 $ at temperature $T/T_c\sim 0.2$
yielding a gap of order $\Delta_2(0)/\Delta_{\rm BCS}\sim 0.2$.
The stronger the interband coupling the less visible the feature of
this transition becomes.
For strong interband coupling all gaps are tied together
and act like a superconductor with a single order parameter.
The ZR fitting of the specific heat suggests that we are in an
intermediate regime of interband coupling, $g_3/g_1=0.07$.

\begin{figure}[hb]
\onefigure{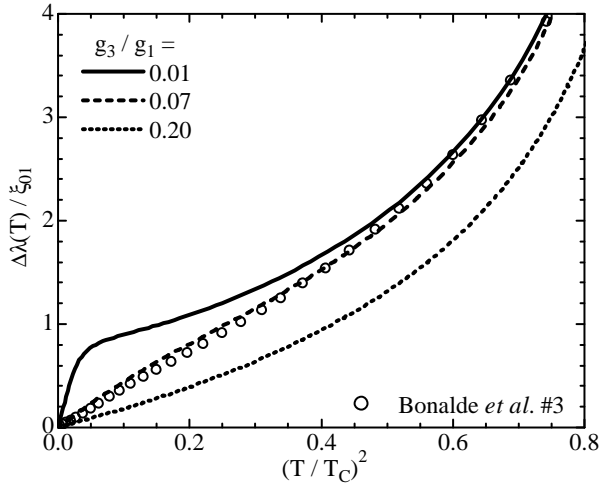}
\caption{\label{fig2} The temperature dependence of the penetration
  depth for the different $g_3/g_1$. The other parameters are fixed as
  $g_1=0.4$, $g_2/g_1=0.85$ and $\kappa_{01}=2.0$. A moderate
  interband coupling fits well the observed temperature dependence.}
\end{figure}
In Fig.~\ref{fig2} we compare the calculated penetration depth for
various values of $ g_3/g_1 $ and  $\kappa_{01}=2.0$ with the
experimental data by Bonalde et al. \cite{Bonalde00} with
$\xi_{01}=700$\AA$\,$ \cite{Riseman98}.
For the weak interband coupling, the penetration depth increases
rapidly at low temperature below $T/T_c\sim0.2$, which reflects both the
smallness of the passive-band gap, $\Delta_2(0)$, and the rapid decrease
of $\Delta_2(T)$ at $T/T_c\sim 0.2$.
For the strong interband coupling, the $T^2$-increase due to
the passive bands is easily overshadowed by the active-band contribution,
showing a monotonous increase with concave shape.
Intermediate interband couplings give a temperature dependence with a
weak convexo-concave shape by subtly
combining the contributions from both bands.
The experimental data indeed exhibit this type of behavior.
The moderate interband coupling, $g_3/g_1=0.07$, which is exactly the
same parameter as ZR to fit the specific heat measurement, also
reproduces the observed temperature dependence over the wide temperature range.

\begin{figure}[th]
\onefigure{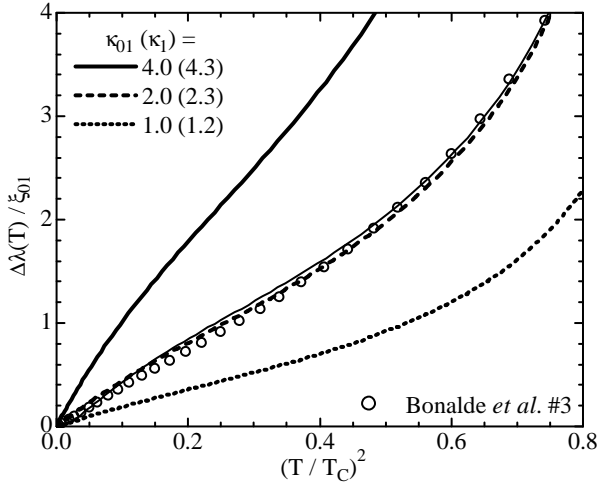}
\caption{\label{fig3} The $\kappa_{01}$ dependence for $g_1=0.4$,
  $g_2/g_1=0.85$ and $g_3/g_1=0.07$. The resultant Ginzburg-Landau
  parameters, $\kappa_1=\lambda(0)/\xi_{01}$ are shown in the
  parenthesis. The thin solid line is for the nodeless gap in passive
  bands (see text in detail).}
\end{figure}
Next we fix the interband coupling constant to the value optimized by ZR
$g_3/g_1=0.07$ and use $\kappa_{01}$ as the only free fitting parameter.
Figure~\ref{fig3} shows curves for various values of $\kappa_{01}$,
where the resultant GL parameter for the active band,
$\kappa_1=\lambda(0)/\xi_{01}$, is shown in the parenthesis.
The slopes of the whole curve monotonously decrease starting from the London
limit and approaching the Pippard limit. Irrespective of a measure of
non-locality, all curves show the convexo-concave shape.
The best fitting for the experimental data is given by
$\kappa_{01}=2.0$ with $\kappa_1=2.3$, which is slightly smaller than
the generally cited value $\kappa\approx2.6$ \cite{Riseman98}.
We also estimate $\kappa_2/\kappa_1=\xi_{01}/\xi_{02}\approx0.66$.
The small discrepancy between measured $ \kappa$ and ours is
very plausible, in view of the fact that $ \kappa = 2.6 $ results
from the measurement of the London penetration depth by means of $
\mu$SR in the mixed state. The order parameter in the passive
bands with a larger coherence length is more easily diminished by a
magnetic field than that of the active band. Consequently, the
$ \mu$SR-measurement observes a larger London penetration depth, where
screening currents from the $ \alpha $-$\beta$-bands are reduced.
This explains the smaller value of the very-low-field penetration depth as measured
by Bonalde et al. \cite{Bonalde00,Riseman98}. Our fit suggests a low-field
value of $ \lambda(0) \approx 1600 $\AA$\,$ compared to $\mu$SR-derived
value of $\lambda(0)\approx1900$\AA$\,$ which is rather likely only due to the $
\gamma $-band contribution \cite{Riseman98}.

So far we have assumed horizontal line nodes in the
passive bands due to interlayer Cooper pairing (ZR scenario).
As an opposite limit, we consider the nodeless gap, ${\bm d}_2={\bm z}\Delta_2(T)(\hat{k_x}+i\hat{k_y})$, corresponding to the case of the intralayer pairing only.
Trying the same fitting procedure we find only a minor difference at very low
temperatures as shown in Fig.~\ref{fig3} (thin solid line).
While the difference is small in $\lambda(T)$, the removal of the nodes would
be more easily visible in the specific heat data.

We have shown that the nonlocal response for the
horizontal line node
yields $-T^2\ln T$ behavior in the London limit, while $T^2$ power law in
the Pippard limit yet at quite low temperature.
On the other hand, the ODS model in combination with nonlocal response naturally explains
the apparent $T^2$-dependence in a quantitatively consistent way with the specific heat fitting.
Concerning the question of nodes, it seems that the London penetration depth data
are less affected by nodes than the specific heat data.
Eventually, a more anisotropic but nodeless gap shape would likely be sufficient
for good fits in any case. In view of the complicated quasiparticle spectrum in the multi-gap compound Sr$_2$RuO$_4$, it is difficult from present experimental data to come to conclusive
answer concerning the detailed gap shape.

\acknowledgments
We thank M. Matsumoto, Y. Maeno, T.M. Rice,
M.E. Zhitomirsky and I. Bonalde for fruitful discussions.
We are indebted to D.J. Van Harlingen for sending experimental
data and useful discussions.
H.K. is supported by a Grant-in-Aid for encouragement of Young
Scientists from Monkasho of Japan.
This work was also supported by Swiss National Fund.


\begin{thebibliography}{0}
\bibitem{Maeno94} Maeno Y. {\it et al}., Nature (London) {\bf 372} (1994) 532.
\bibitem{Maeno01} See for example, Maeno Y., Rice T.M. and Sigrist M.,
  Phys. Today {\bf 54} No. 1 (2001) 42.
\bibitem{Rice95} Rice T.M. and Sigrist M., J. Phys. Condens. Matter
  {\bf 7} (1995) L643; Baskaran G., Physica B {\bf 223-224} (1996) 490.
\bibitem{Ishida98} Ishida K. {\it et al}., Nature (London) {\bf 396} (1998) 658.
\bibitem{Luke98} Luke G.M. {\it et al}., Nature (London) {\bf 394} (1998) 558.
\bibitem{Hasegawa00} Hasegawa Y., Machida K. and Ozaki M.,
  J. Phys. Soc. Jpn. {\bf 69} (2000) 336.
\bibitem{Sigrist99} Sigrist M. {\it et al}., Physica C{\bf 317} (1999) 134.
\bibitem{Mackenzie96} Mackenzie A.P. {\it et al}.,
  Phys. Rev. Lett. {\bf 76} (1996) 3786.
\bibitem{NishiZaki9900} NishiZaki S., Maeno Y. and Mao Z., J. Low
  Temp. Phys. {\bf 117} (1999) 1581; J. Phys. Soc. Jpn. {\bf 69} (2000) 572.
\bibitem{Ishida00} Ishida K. {\it et al}., Phys. Rev. Lett. {\bf 84} (2000) 5387.
\bibitem{Tanatar01} Tanatar M. {\it et al}., Phys. Rev. Lett. {\bf 86}
  (2001) 2649; Phys. Rev. B{\bf 63} (2001) 064505.
\bibitem{Izawa01} Izawa K. {\it et al}., Phys. Rev. Lett. {\bf 86} (2001) 2653.
\bibitem{Lupien01} Lupien C. {\it et al}., Phys. Rev. Lett. {\bf 86} (2001) 5986.
\bibitem{Miyake99} Miyake K. and Narikiyo O., Phys. Rev. Lett. {\bf
    83} (1999) 1423.
\bibitem{Graf00} Graf M.J. and Balatsky A.V., Phys. Rev. B{\bf 62} (2000) 9697.
\bibitem{Dahm00} Dahm T., Won H. and Maki K., cond-mat/0006301 (unpublished).
\bibitem{Zhitomirsky01} Zhitomirsky M.E. and Rice T.M.,
  Phys. Rev. Lett. {\bf 87} (2001) 057001.
\bibitem{Agterberg97} Agterberg D.F., Rice T.M. and Sigrist M.,
  Phys. Rev. Lett. {\bf 78} (1997) 3374.
\bibitem{Imai98} Imai T. {\it et al}., Phys. Rev. Lett. {\bf 81}
  (1998) 3006.
\bibitem{Ng01} Ng K.K. and Sigrist M., J. Phys. Soc. Jpn. {bf 69}
  (2000) 3764.
\bibitem{Mazin9799} Mazin I.I. and Singh D.J., Phys. Rev. Lett. {\bf
    79} (1997) 733; {\bf 82} (1999) 4324.
\bibitem{Sidis99} Sidis Y., Braden M., Bourges P., Hennion B.,
  NishiZaki S., Maeno Y. and Mori Y., Phys. Rev. Lett. {\bf 83} (1999)
    3320.
\bibitem{Ng00} Ng K.K. and Sigrist M., Europhys. Lett. {\bf 49} (2000) 473.
\bibitem{Blount85} Blount E.I., Phys. Rev. B{\bf 32} (1985) 2935.
\bibitem{Bonalde00} Bonalde I. {\it et al}., Phys. Rev. Lett. {\bf 85}
  (2000) 4775.
\bibitem{Kosztin97} Kosztin I. and Leggett A.J., Phys. Rev. Lett. {\bf
    79} (1997) 135.
\bibitem{Morinari00} Morinari T. and Sigrist M.,
  J. Phys. Soc. Jpn. {\bf 69} (2000) 2411.
\bibitem{Kubo00} Kubo K. and Hirashima D.S., J. Phys. Soc. Jpn. {\bf
    69} (2000) 3489.
\bibitem{Riseman98} Riseman T.M., {\it et al}., Nature (London) {\bf 396} (1998) 242.
\end{thebibliography}
\end{document}